\documentclass[aps,prl,preprint]{revtex4-1}
\usepackage{graphicx}

\begin{document}


\title{A Mechanism of Ultraviolet Naturalness}


\author{Durmu{\c s} Demir}
\email[]{demir@physics.iztech.edu.tr}
\homepage[]{http://physics.iyte.edu.tr/staff/durmus-ali-demir/}
\affiliation{Department of Physics, Izmir Institute of Technology, 35430, Izmir, Turkey}


\date{\today}

\begin{abstract}
The Standard Model (SM), as the quantum field theory of the strong
and electroweak interactions, needs be carried into curved spacetime 
to incorporate gravity. This is done here not for the full but for the
effective SM action by integrating-in affine curvature. This approach
leads to naturalization of the SM Higgs sector and reduction of the
cosmological constant down to the neutrino scale thanks to the 
fact that quadratic (quartic) UV contributions to the Higgs boson mass (vacuum energy) 
turn into the Higgs-curvature coupling (the Einstein-Hilbert term). 
New physics beyond the SM, necessary for inducing gravity properly, does
not have to interact with the SM. It can form a secluded sector to 
source non-interacting dark matter observable via only gravity, 
or a weakly-coupled sector to source dark matter and possible collider 
signals. 
\end{abstract}


\maketitle



The SM, experimentally verified to describe the physics at the Fermi scale ${\textstyle{G_F}}$,
develops the effective action
\begin{eqnarray}
\label{action-SM-flat}
S^{eff}_{\Lambda}(\eta) = S_{G_F}\left({\eta}_{\mu\nu},
\log\left( G_F \Lambda^2\right)\right) + S_{\Lambda}\left({\eta}\right)
\end{eqnarray}
after integrating out quantum field fluctuations ranging in
frequency from ${\textstyle{G_F^{-1/2}}}$ up to an ultraviolet (UV)
scale ${\textstyle{\Lambda}}$, which can be as high as the
gravitational scale ${\textstyle{M_{Pl}}}$. Here, $\Lambda$  is a
physical scale marking the start of the UV physics that completes
the SM.

The first part of (\ref{action-SM-flat}),
$S_{G_F}\left({\eta}_{\mu\nu}, \log\left( G_F
\Lambda^2\right)\right)$, involves the tree-level SM interactions
improved by logarithmic UV contributions, where the flat metric
${\textstyle{\eta_{\mu\nu}}}$ is there to enable kinetic terms. In
its ground state, the Higgs doublet $H$, endowed with the
condensation parameter ${\textstyle{{\textstyle{m_H^2 \simeq -
G^{-1}_{F}}}}}$, condenses as ${\textstyle{\langle H^{\dagger}
H\rangle \simeq G^{-1}_F}}$ to generate particle masses in units of
${\textstyle{G^{-1/2}_{F}}}$ and to deposit a vacuum energy of size
${\textstyle{G^{-2}_{F}}}$. The net vacuum energy can always be
zeroed \begin{eqnarray} \label{nat-vac} \langle
S_{G_F}\left({\eta}_{\mu\nu}, \log\left( G_F
\Lambda^2\right)\right)\rangle = 0
\end{eqnarray} by suitably choosing its incalculable tree-level part.
This ground state, the electroweak vacuum, is phenomenologically
viable and technically natural \cite{natural}.

The second part of (\ref{action-SM-flat})
\begin{eqnarray}
\label{action-UV} S_{\Lambda}\left(\eta\right)= \int d^4x \sqrt{
\left\Vert\eta\right\Vert} \Big\{ c_4 \Lambda^4 + c_2 m_H^2
\Lambda^2 + c_H \Lambda^2 H^{\dagger} H \Big\}
\end{eqnarray}
encodes quantum corrections to vacuum and Higgs sectors. It is
power-law in the UV scale $\Lambda$. It can be small or large
depending on the Higgs field and the UV scale. For a UV-sized
$\langle H^{\dagger} H\rangle$, which is physically unacceptable, it
is possible to suppress $S_{\Lambda}\left(\eta\right)$. For
${{\langle H^{\dagger} H\rangle \simeq G^{-1}_F}}$, as in the
physical electroweak vacuum, however, the same action can be
hierarchically large depending on how large $\Lambda$ is. In fact,
the correction ${{c_H \Lambda^2}} \cong - \left(\Lambda/4\right)^2$
to Higgs boson mass exceeds is already at ${{\Lambda\cong 550\ {\rm
GeV}}}$. This means that the power-law UV contributions in
$S_{\Lambda}\left(\eta\right)$ render the SM unnatural just above
the Fermi scale if the Higgs field is in the electroweak vacuum,
that is, in the ground state of $S_{G_F}\left({\eta}_{\mu\nu},
\log\left( G_F \Lambda^2\right)\right)$ \cite{natural}. It is due to
this unnaturalness that the natural extensions of the SM
(supersymmetry, extra dimensions, technicolor and their derivatives)
have been expected to show up at the Fermi scale, more precisely,
around $550\ {\rm GeV}$. They have not, however, even glimpsed in
the LHC searches which now probe ${\rm TeV}$ energies
\cite{higgs,lhc-search}. Thus, the SM must be utilizing a different
mechanism which:
\begin{enumerate}
\item must fix ${\textstyle{\Lambda}}$ to a phenomenal scale
for loop integrations to be {{physical}} rather than arbitrary
regularizations,

\item must necessitate no new visible particles in charted
domains to be compatible with the LHC results, and

\item must involve no fine-tunings to qualify {{natural}}.
\end{enumerate}
These requirements seem to block all roads but {{gravity}}. Yet, the
SM continues to be unnatural in curved spacetime. It can be
stabilized at the Fermi scale with a severe yet harmless fine-tuning
in the Higgs-curvature coupling \cite{demir3}. The same coupling
facilitates stabilization via also nonlinear dynamics above the
Fermi scale \cite{bij}. It is unnatural also in Sakharov's induced
gravity where $M_{Pl} \propto \Lambda$ \cite{sakharov}. It can,
however, go natural if gravity is generated by {{integrating-in
curvature}} upon the SM effective action in flat spacetime. Indeed,
in the ground state of ${{S_{G_F}\left({\eta}_{\mu\nu}\right)}}$
where (\ref{nat-vac}) holds, the total action  $\langle
S^{eff}_{\Lambda}(\eta) \rangle$ reduces to $\langle
S_{\Lambda}\left(\eta\right)\rangle$ and constancy of the action
density facilitates introduction of the affine action
\begin{eqnarray}
\label{volume-affine}
\langle S^{eff}_{\Lambda}(R) \rangle = \langle S_{\Lambda}(R) \rangle
= \int d^4x \sqrt{ \left\Vert\frac{{{R\left(\Gamma\right)}}}{\Lambda^{2}}\right\Vert}\Big\{ c_4
\Lambda^{4} + c_2 m_H^2 \Lambda^{2} + c_H \Lambda^{2} \langle H^{\dagger} H\rangle  \Big\}\;\;\;
\end{eqnarray}
where ${{\Gamma^{\lambda}_{\mu\nu}}}$ is the affine connection and
${\textstyle{{{R}}_{\mu\nu}\left(\Gamma\right)}}$ is its Ricci
tensor \cite{eddington,demir1}. This metamorphosis of $\langle
S_{\Lambda}\left(\eta\right)\rangle$ into $\langle S_{\Lambda}(R)
\rangle$ is justified by the fact that $\langle
S_{\Lambda}\left(\eta\right)\rangle$ is nothing but the stationary
value of $\langle S_{\Lambda}(R) \rangle$. Indeed, $\langle
S_{\Lambda}(R) \rangle$ stays stationary if affine curvature obeys
the equation of motion
\begin{eqnarray} \label{eom} \nabla^{\Gamma}_{\alpha}
\left( \sqrt{\left\Vert
\frac{R\left(\Gamma\right)}{\Lambda^2}\right\Vert}
\left\{\left(\frac{R\left(\Gamma\right)}{\Lambda^2}\right)^{-1}\right\}^{\mu\nu}\right)
= 0 \end{eqnarray} whose solution (the so-called Eddington solution
\cite{eddington,demir1})
\begin{eqnarray} \label{soln}
R_{\mu\nu}\left(^{{\bar{g}}}\Gamma\right) = \Lambda^2
{\bar{g}}_{\mu\nu} \end{eqnarray} reduces $\langle S_{\Lambda}(R)
\rangle$ to $\langle S_{\Lambda}({\bar{g}}) \rangle$. Here, ${\bar
g}_{\mu\nu}$ is a covariantly-constant tensor that acts as metric
tensor \cite{demir1} with Levi-Civita connection
${\textstyle{^{{\bar{g}}}\Gamma^{\lambda}_{\mu\nu}}}$ and Ricci
tensor ${\textstyle{R_{\mu\nu}\left(^{{\bar{g}}}\Gamma\right)}}$. It
relaxes to the flat metric, ${\bar g}_{\mu\nu} \rightarrow
\eta_{\mu\nu}$, in the absence of curvature terms.

It should be clear that, the reduction of $\langle S_{\Lambda}(R)
\rangle$ into $\langle S_{\Lambda}({\bar{g}}) \rangle$ means
integration of the spacetime curvature out of dynamics. Speaking
conversely, affine curvature can be integrated in $\langle
S_{\Lambda}\left(\eta\right)\rangle$ to carry it into curved
spacetime. This observation reveals a method to map the SM effective
action in flat spacetime into curved affine spacetime, where UV
sensitivity can get modified by curvature effects. This because the
quartic and quadratc $\Lambda$ terms in (\ref{volume-affine}) get
modified by the $1/\Lambda$ factor in the determinant.

It is clear that the metamorphosis in (\ref{volume-affine}) holds in
the ground state of
${\textstyle{S_{G_F}\left({\eta}_{\mu\nu}\right)}}$. Perturbations
about this vacuum state affects the affine dynamics. The least
affected piece is $S_{\Lambda}\left(R\right)$ because replacement of
${\textstyle{\langle H^{\dagger} H\rangle}}$ with ${\textstyle{
H^{\dagger} H}}$ does not modify (\ref{volume-affine}) in form. This
is not the case for
${\textstyle{S_{G_F}\left({\eta}_{\mu\nu}\right)}}$, however. The
reason is that it involves metric tensor -- an object having no
place in affine geometry. As a matter of fact, for
${\textstyle{S_{\Lambda}\left(R\right)}}$ in affine geometry and
${\textstyle{S_{G_F}\left(g_{\mu\nu}\right)}}$ in metrical geometry
to amalgamate to form the SM action in curved geometry,
${\textstyle{R_{\mu\nu}\left(\Gamma\right)}}$ and
${\textstyle{g_{\mu\nu}}}$ must be bridged by an auxiliary relation.
The requisite relation, a {{constitutive law}} bridging the two
geometries, must reduce, when
${\textstyle{S_{G_F}\left(g_{\mu\nu}\right)}}\rightarrow 0$, to the
motion equation (\ref{soln}) because it is the definition of metric.
In consequence, the specific relationship \begin{eqnarray}
\label{complement} R_{\mu\nu}\left(\Gamma\right) = \frac{1}{2}
\sqrt{R\left(g,^{g}\Gamma\right)} \Lambda g_{\mu\nu} \end{eqnarray}
qualifies as a proper constitutive law because it tends to
(\ref{soln}) as ${\textstyle{S_{G_F}\left(g_{\mu\nu}\right)}}
\rightarrow 0$. In this very limit, $g_{\mu\nu} \rightarrow
{\bar{g}}_{\mu\nu}$ and hence
${\textstyle{R\left({{g}},^{{{g}}}\Gamma\right) \rightarrow
R\left({\bar{g}},^{{\bar{g}}}\Gamma\right) \equiv
({\bar{g}}^{-1})^{\mu\nu}
R_{\nu\mu}\left(^{{\bar{g}}}\Gamma\right)}}$ with
${\textstyle{R\left({\bar{g}},^{{\bar{g}}}\Gamma\right) = 4
\Lambda^2}}$. Thus, with the constitutive law (\ref{complement}),
$S_{\Lambda}\left(R\right)$ transforms into \begin{eqnarray}
\label{UV-part-tildetilde} S_{\Lambda}\left(g\right)= \int d^4x
\sqrt{ \left\Vert g\right\Vert}\left( c_4 \Lambda^{2} + c_2
m_{H}^{2}  + c_H H^{\dagger}
H\right)R\left(g,^{g}\Gamma\right)\;\;\; \end{eqnarray} which hints
at the Einstein-Hilbert action, but already at one loop
\begin{eqnarray} c_4 = \frac{1}{64 \pi^2} (n_b-n_f) \end{eqnarray}
is negative for $n_b = 28$ bosonic and $n_f = 90$ fermionic degrees
of freedom in the SM. This means that, gravity can be generated as
an attractive force only if the SM is extended by {{new physics}}
(NP) having a spectrum ${{n}}^{{{NP}}}_b - {{n}}^{{{NP}}}_f \geq 63$
at a scale $\Lambda_{{{NP}}}$. This ensures only attractive nature
of gravity. There is more to it in that the UV scale of any field
theory must be bounded by the fundamental scale of gravity. In other
words, it is necessary to have $\Lambda \leq M_{Pl}$. Imposing this
leads to the condition
\begin{eqnarray}
\label{sart}
{{n}}^{{{NP}}}_b - {{n}}^{{{NP}}}_f \gtrsim 128 \pi^2 + 62 \simeq
1325
\end{eqnarray}
which means that the NP is a rather crowded mostly-bosonic sector.
It is clear that these extra fields do not, partly or wholly, have
to interact with the SM spectrum. All they are required is to
provide the excess bosonic degrees of freedom necessary to induce
gravity correctly in sign and in strength. This NP develops the UV
action
 \begin{eqnarray} \label{action-UV-NP}
{{S}}^{{\text{NP}}}_{\Lambda}\left(\eta\right)=\int d^4 x \sqrt{
\left\Vert\eta\right\Vert}\, {{c}_{4}^{{\text{NP}}}} \Lambda^{4}
\end{eqnarray} if its spectrum contains no scalar fields. This
action, with ${{c}_{4}^{{\text{NP}}}} = \left({{n}}^{{{NP}}}_b -
{{n}}^{{{NP}}}_f\right)/64\pi^2$ as follows from (\ref{sart}), gives
a gravity sector like (\ref{UV-part-tildetilde}) after
integrating-in curvature as in (\ref{volume-affine}) so that the
combined ${{SM+NP}}$ action \begin{eqnarray}
\label{action-Fermi-curve} {{S}}_{G_{F}}\left(g_{\mu\nu},
\log\left(G_F \Lambda^{2}\right)\right) +
{{S}}_{\Lambda_{{\text{NP}}}}\left(g_{\mu\nu},
\log\left(\Lambda^{2}/\Lambda_{{\text{NP}}}^{2}\right)\right) +
S_{G}(g) \;\;\;\;\;\;\; \end{eqnarray} in curved spacetime possesses
the gravity sector \begin{eqnarray} \label{Hilbert}
S_{G}\left(g\right) = \int d^{4}x \sqrt{\Vert g \Vert} \left(
\frac{1}{2} M_{Pl}^2 + \zeta_{H} H^{\dagger} H
\right)R\left(g,^{g}\Gamma\right)\;\;\;\; \end{eqnarray} wherein the
curvature couplings $\zeta_H = c_H/4$ and \begin{eqnarray}
\label{params} M_{Pl}^2 &=& \frac{1}{2}\left(\left(c_4 +
{c}_4^{{{NP}}}\right) \Lambda^2 + c_2 m_H^2 \right) \end{eqnarray}
are entirely fixed by the flat spacetime physics. Gravity gets
induced correctly thanks to the NP contribution in (\ref{sart}).

It must be clear from (\ref{Hilbert}) that, the sources of
ultraviolet unnaturalness are completely defused: The {{UV-sized}}
{{vacuum energy sources the gravitational constant}} in lieu of the
cosmological constant (as was proposed first in
\cite{demir2,demir4}), and the {{UV-sized Higgs boson mass converts
into}} {{nonminimal Higgs-curvature coupling $\zeta_{H}$}}. These
transmuted interactions ensure the stabilization of the SM at the
Fermi scale \cite{demir3}. The end result is that the unnatural SM
in flat spacetime is metamorphosed into the natural SM in curved
spacetime in collaboration with an NP sector. In other words,
naturalness in the SM is achieved by carrying not the full but the
effective SM action into curved spacetime. The latter is clearly
more harmonious with the classical curved geometry.

Constitutive laws set gravity by specifying the spacetime geometry
just as they do electrodynamics by specifying the material
permittivity and permeability. They are vital for dynamical
completeness. The constitutive law in (\ref{complement}) leads to
the Einstein gravity. Its extension
${\textstyle{R_{\mu\nu}\left(\Gamma\right) = {1}/{4} \big(
\sqrt{R\left(g,^{g}\Gamma\right)} \Lambda + {1}/{2}
R\left(g,^{g}\Gamma\right) \big) g_{\mu\nu}}}$ gives
higher-curvature gravity with ${\textstyle{\Lambda
R^{3/2}\left(g,^{g}\Gamma\right)}}$ and
${\textstyle{R^2\left(g,^{g}\Gamma\right)}}$ terms to be contrasted
with Sakharov's induced gravity \cite{sakharov}. Its simpler forms,
${\textstyle{R_{\mu\nu}\left(\Gamma\right) = {1}/{4}
R\left(g,^{g}\Gamma\right) g_{\mu\nu}}}$ and
${\textstyle{R_{\mu\nu}\left(\Gamma\right) = \Lambda^2
g_{\mu\nu}}}$, lead, respectively, to quadratic curvature and zero
curvature theories. These examples, all adhering to (\ref{soln})
when ${\textstyle{S_{G_F}\left(g_{\mu\nu}\right)}}$ and
${\textstyle{S_{\Lambda_{{\text{NP}}}}\left(g_{\mu\nu}\right)}}$ are
absent, ensure that
${\textstyle{\sqrt{R\left(g,^{g}\Gamma\right)}}}$ is essential for
Einstein gravity.

Concerning the construction of (\ref{volume-affine}), one may wonder
why the curvature integrated in is affine but not metrical. The
reason for this is that one then obtains the action
${\textstyle{\int d^4x \sqrt{ \left\Vert
g\right\Vert}\Big\{\frac{1}{2} M_{Pl}^2 R\left(g,^{g}\Gamma\right)
-c_4 \Lambda^4 - c_2 m_H^2 \Lambda^2 - c_H \Lambda^2 H^{\dagger} H
\Big\}}}$ instead of ${\textstyle{S_{\Lambda}\left(g\right)}}$ in
(\ref{UV-part-tildetilde}). This alters the sign of vacuum energy.
Also, ${\textstyle{S_{G_F}\left(g_{\mu\nu}\right)}}$ develops
anomalous interactions via the trace of the energy-momentum tensor.
Therefore, {{integrating-in GR causes unphysical effects}} plaguing
the curved spacetime field theory.

Comparatively eying, quartic and quadratic UV contributions, while
cancel out in SUSY, convert into gravity in ${{SM+NP}}$. In SUSY,
suppressing the Higgs mass shift
\begin{eqnarray}
\delta m_h^2 \propto
\Lambda_{{{SUSY}}}^2
\log\left(\Lambda^2/\Lambda^2_{{{SUSY}}}\right)
\end{eqnarray}
requires ${{\Lambda^2_{{{SUSY}}} \cong G_F^{-1}}}$, which must be
in tension with LHC if not with naturalness. In ${{SM+NP}}$, this
logarithmic unnaturalness does not have to arise simply because SM
and NP do not have to interact non-gravitationally for the mechanism
to work. Indeed, NP can form a completely secluded sector to have
only  gravitational interactions with the SM, as assumed in deriving
(\ref{Hilbert}). This NP can source a non-interacting dark matter
which reveals itself via only its weight through flat rotation
curves of galaxies \cite{dark}, for instance. This dark matter,
impossible to observe in direct searches, can be conveniently called
as pitch-dark or ebony matter. To this end, it is noteworthy that
the latest direct search results seem to increasingly disfavor
weakly-interacting massive particles at the Fermi scale as dark
matter candidates \cite{dark-exp}.

Contrarily, if NP is weakly-coupled then one is led to the usual
dark matter and weak signals at the LHC \cite{lhc-search}. If
coupling is significant then ${\textstyle{\zeta_H \rightarrow
\zeta_H + c_H^{{{NP}}}/4}}$ in (\ref{Hilbert}) and
${\textstyle{{{S}}_{\Lambda_{{{NP}}}}\left(g_{\mu\nu},
\log\left(\Lambda^{2}/\Lambda_{{{NP}}}^{2}\right)\right)}}$ in
(\ref{action-Fermi-curve}) shifts Higgs mass by
\begin{eqnarray}
\delta m_h^2  \propto \Lambda_{{{NP}}}^2
\log\left(\Lambda^2/\Lambda^2_{{{NP}}}\right)
\end{eqnarray}
that is waned if ${{\Lambda_{{{NP}}} \cong G_{F}^{-1/2}}}$. This causes
logarithmic unnaturalness, as in SUSY \cite{martin}. The searches at the LHC,
which have now reached energies fairly above the Fermi scale, have 
found no significant signal of any new scalar field. In this 
sense, it would not be unrealistic to anticipate that the NP 
may not involve any scalar fields.

Here is the summary. The SM is the quantum field theory of the
strong and electroweak interactions. To include gravity as the
fourth force it is necessary to carry the SM into curved spacetime.
This is done in the present work by taking not the full but the
effective SM action into curved spacetime. This is more natural in
view of the harmony between the quantized matter with effective
dynamics and classical curved geometry. Here, carriage into curved
spacetime is done by integrating-in affine curvature. This leads to
eradication of the Higgs naturalness problem (Higgs mass is of order
$G_{F}^{-1/2}$), suppression of the cosmological constant towards
the neutrino scale (vacuum energy is at most of order $G_{F}^{-2}$),
and formation of a viable dark matter candidate (NP can be
secluded). The workings of the mechanism is such that gravity, not
included in the spectrum of forces in the $SM$, is brought in by
incorporating it as a means for naturalizing the model. Further work
is needed for modeling the NP, elucidating dark matter, downsizing
cosmological constant to Hubble scale \cite{demir1}, explicating
gravity sector, and studying various concurrent phenomena.

The author thanks I. Antoniadis and R. G{\"u}ven for conversations on integrating out gravity, G. Demir for discussions on electromagnetic constitutive laws, H. Azri and K. G{\"u}ltekin for conversations on affine gravity,
C. Karahan and B. Korutlu for discussions on naturalness, and O. Do{\~g}ang{\"u}n, A. {\"O}zpineci and {\.I}. Turan for reading the manuscript and conversations.  This work is supported in part by the T{\"U}B{\.I}TAK grant
115F212.

\end{document}